\documentclass[showpacs,preprintnumbers,amsmath,amssymb,superscriptaddress]{revtex4}

\usepackage{graphicx}
\usepackage{bbm}
\usepackage{bm}

\begin{document}


\title{Surface-wave solitons on the interface between a linear medium and a nonlocal nonlinear  medium}

\author{Zhiwei Shi}
\affiliation{Laboratory of Photonic
Information Technology, South China Normal University, Guangzhou
510631, P.R.China}
\affiliation{Faculty of Information Engineering, Guangdong University of Technology.}
\author{Huagang Li}
\affiliation {Department of Physics, Guangdong Education Institute,
Guangzhou, 510303, P. R. China}%

\author{Qi Guo}%
\email{guoq@scnu.edu.cn} \affiliation{Laboratory of Photonic
Information Technology, South China Normal University, Guangzhou
510631, P.R.China}

\date{\today}

\begin{abstract}
We address the properties of surface-wave solitons on the interface
between a semi-infinite homogeneous linear medium and a
semi-infinite homogeneous nonlinear nonlocal medium. The stability,
energy flow and FWHM of the surface wave solitons can be affected by
the degree of nonlocality of the nonlinear medium. We find that the
refractive index difference affects the power distribution of the
surface solitons in two media. We show that the different boundary
values at the interface can lead to the different peak position of
the surface solitons, but it can not influence the solitons
stability with a certain degree of nonlocality.
\end{abstract}
\maketitle

Surface waves propagating along the interface between a homogeneous
linear medium and a homogeneous nonlinear medium display many
interesting properties, which have no analogues in homogeneous
media. Decades years ago, such surface waves had been studied in the
local nonlinear optical case~\cite{ref1,ref2,ref3,ref4,ref5,ref6}.
[1] and [6] show that there is no stable surface wave when the zero
field refractive index of the nonlinear medium is larger than the
refractive index of the linear medium, on the contrary, surface wave
is stable.

In the nonlocal nonlinear optical domain, such surface waves were
analyzed at the interfaces of diffusive Kerr-type
materials~\cite{ref7,ref8,ref9}. Recently, surface-wave solitons
were observed at the interface between a dielectric medium(air) and
a nonlocal nonlinear medium(lead glasses)~\cite{ref10}. They found
that these solitons are always attracted toward the surface, and
unlike their Kerr-like counterparts, they do not exhibit a power
threshold. Two-dimensional surface solitons featuring topologically
complex shapes, including vortices and dipoles with nodal lines
perpendicular to the interface of nonlocal thermal media were
studied in [11]. Defocusing thermal materials can also support
surface waves under appropriate conditions~\cite{ref12,ref13}.
Multiploe solitons localized at a thermally insulating interface are
addressed in [14].

However, to our knowledge, the variation of such surface-wave
solitons due to the change of the degree of nonlocality or the
boundary value at the interface of the semi-infinite nonlocal
nonlinear media and the semi-infinite linear media were not studied
to this day. In this Letter, we reveal that the degree of
nonlocality can affect the stability, the energy flow and the full
width at half maximum(FWHM) of the surface solitons. We state that
the refractive index difference affects the power distribution of
the surface solitons in two media. In addition, we show that the
different boundary values at the interface can lead to the different
peak position of the surface solitons, but it can not influence the
solitons stability with a certain degree of nonlocality.

Here, we consider the simple (1+1)D case. $x=0$ is the interface of
the nonlinear medium(a linear refractive index $n_{L}$) and a linear
medium of refractive index $n_{0}$. To describe the propagation of
light beams along $Z$ axis near the interface of the nonlinear
medium, we use a nonlinear Schr\"{o}inger equation for the
dimensionless amplitude $a$ of the light field coupled to the
equation for normalized nonlinear induced change of the refractive
index $\phi$,

\begin{subequations}
\label{one}
\begin{equation}
i\partial_{Z}a+\frac{1}{2}\frac{\partial^{2}a}{\partial
X^{2}}+\frac{w_{0}^{2}}{2}(k_{0}^{2}n_{L}^{2}-\beta^{2})a+\phi
a=0,\textrm{for} -\infty\leq X\leq 0,\label{1a}
\end{equation}
\begin{equation}
i\partial_{Z}a+\frac{1}{2}\frac{\partial^{2}a}{\partial
X^{2}}+\frac{w_{0}^{2}}{2}(k_{0}^{2}n_{0}^{2}-\beta^{2})a=0 \quad
\textrm{for}\quad 0\leq X\leq \infty.\label{1b}
\end{equation}
\end{subequations}
and
\begin{eqnarray}
\label{two} \alpha^{2}\nabla_{\bot}^{2}\phi-\phi+|a|^{2}=0 \quad
\textrm{for} -\infty\leq X\leq 0.\label{2}
\end{eqnarray}
where $X$ is is the transverse coordinate,
$\alpha^{2}=w_{m}^{2}/w_{0}^{2}$, $w_{0}$ is the beam width, $w_{m}$
is the characteristic length of the nonlinear response and $\alpha$
stands for the degree of nonlocality of the nonlinear response.
$\beta$ and $k_{0}=2\pi/\lambda$ are the wave numbers in the media
and in vacuum. $\phi$ is given by
$\phi(X)=\phi_{0}e^{X/\alpha}-(1/\alpha^{2})\int^{0}_{-\infty}G(X,\xi)|a(\xi)|^{2}d\xi$,
where $G(X)=\alpha e^{\xi/\alpha}(e^{X/\alpha}-e^{-X/\alpha})/2$,
for $X\geq\xi$, and $G(X)=\alpha
e^{X/\alpha}(e^{\xi/\alpha}-e^{-\xi/\alpha})/2$, for $X\leq\xi$.
Here, we can safely assume that the boundary condition at the
interface($X=0$) is $\phi=\phi_{0}(\phi_{0}>0)$, where $\phi_{0}=50$
is the initial value, unless we indicate it. $a$ and $\phi$ vanish
at the $X\rightarrow \pm\infty$.

We search for stationary soliton solutions of Eqs.~(\ref{one}) and
(\ref{two}) numerically in the form $a(X,Y,Z)=u(X,Y)\exp(i b Z)$,
where $u$ is the real function and $b$ is a real propagation
constant of spatial solitons in the normalized system.

\begin{subequations}
\label{three}
\begin{equation}
\frac{1}{2}\nabla_{\bot}^{2}u-bu+\frac{w_{0}^{2}}{2}(k_{0}^{2}n_{L}^{2}-\beta^{2})u+\phi
u=0,\label{subeq:3a}
\end{equation}
\begin{equation}
\frac{1}{2}\nabla_{\bot}^{2}u-bu+\frac{w_{0}^{2}}{2}(k_{0}^{2}n_{0}^{2}-\beta^{2})u=0.\label{subeq:db}
\end{equation}
\end{subequations}
and
\begin{eqnarray}\label{eq:four}
\alpha^{2}\nabla_{\bot}^{2}\phi-\phi+|u|^{2}=0.
\end{eqnarray}

To elucidate the linear stability of the solitons, we searched for
perturbed solutions in the form $a(X,Z)=[u(X)+p(X,Z)+iq(X,Z)]\exp(i
b Z)$~\cite{ref9,ref12,ref14,ref15}, where the real $p(X,Z)$ and
imaginary $q(X,Z)$ parts of the perturbation can grow with a complex
rate $\delta=\delta_{r}+i\delta_{i}$ upon propagation. Linearization
of Eq.(3)and (4) around a stationary solution yields the eigenvalue
problem

\begin{subequations}
\label{three}
\begin{equation}
\delta p=-\frac{1}{2}\frac{\partial^{2}q}{\partial X^{2}}+bq-\phi
q-\frac{w_{0}^{2}}{2}(k_{0}^{2}n_{L}^{2}-\beta^{2})q,\label{3a}
\end{equation}
\begin{equation}
\delta q=\frac{1}{2}\frac{\partial^{2}p}{\partial X^{2}}-bp+\phi
p+\frac{w_{0}^{2}}{2}(k_{0}^{2}n_{L}^{2}-\beta^{2})p-u\Delta
\phi.\label{3b}
\end{equation}
\end{subequations}
which holds for $-\infty\leq X\leq 0$. where
$\Delta\phi=2\int^{0}_{-\infty}G(X,\xi)u(\xi)p(\xi)d\xi$.

For $0\leq X\leq \infty$, the eigenvalue problem
\begin{subequations}
\label{four}
\begin{equation}
\delta p=-\frac{1}{2}\frac{\partial^{2}q}{\partial
X^{2}}+bq-\frac{w_{0}^{2}}{2}(k_{0}^{2}n_{0}^{2}-\beta^{2})q,\label{4a}
\end{equation}
\begin{equation}
\delta q=\frac{1}{2}\frac{\partial^{2}p}{\partial
X^{2}}-bp+\frac{w_{0}^{2}}{2}(k_{0}^{2}n_{0}^{2}-\beta^{2})q.\label{4b}
\end{equation}
\end{subequations}

\begin{figure}
\centering
\includegraphics[width=6.0cm]{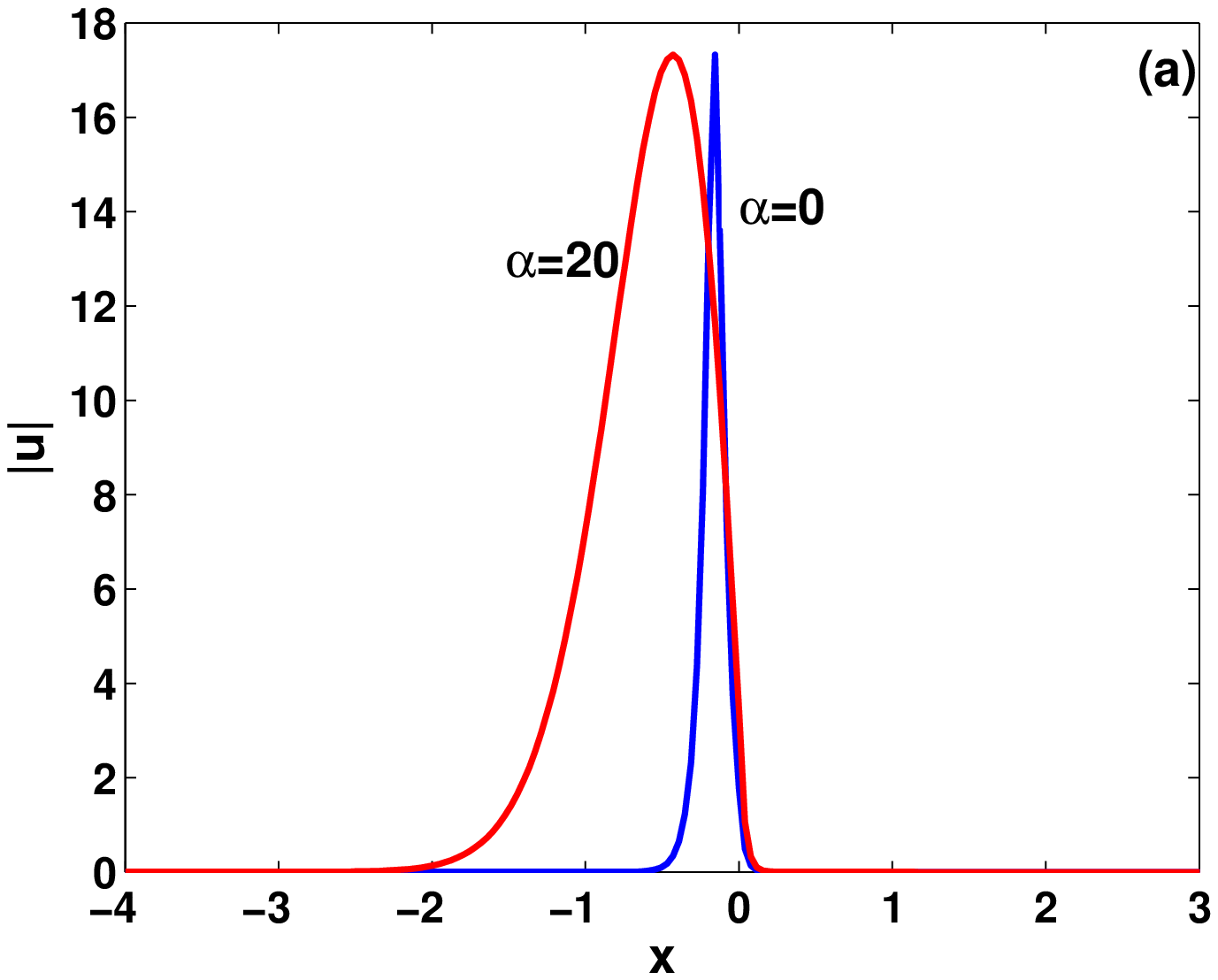}
\includegraphics[width=6.0cm]{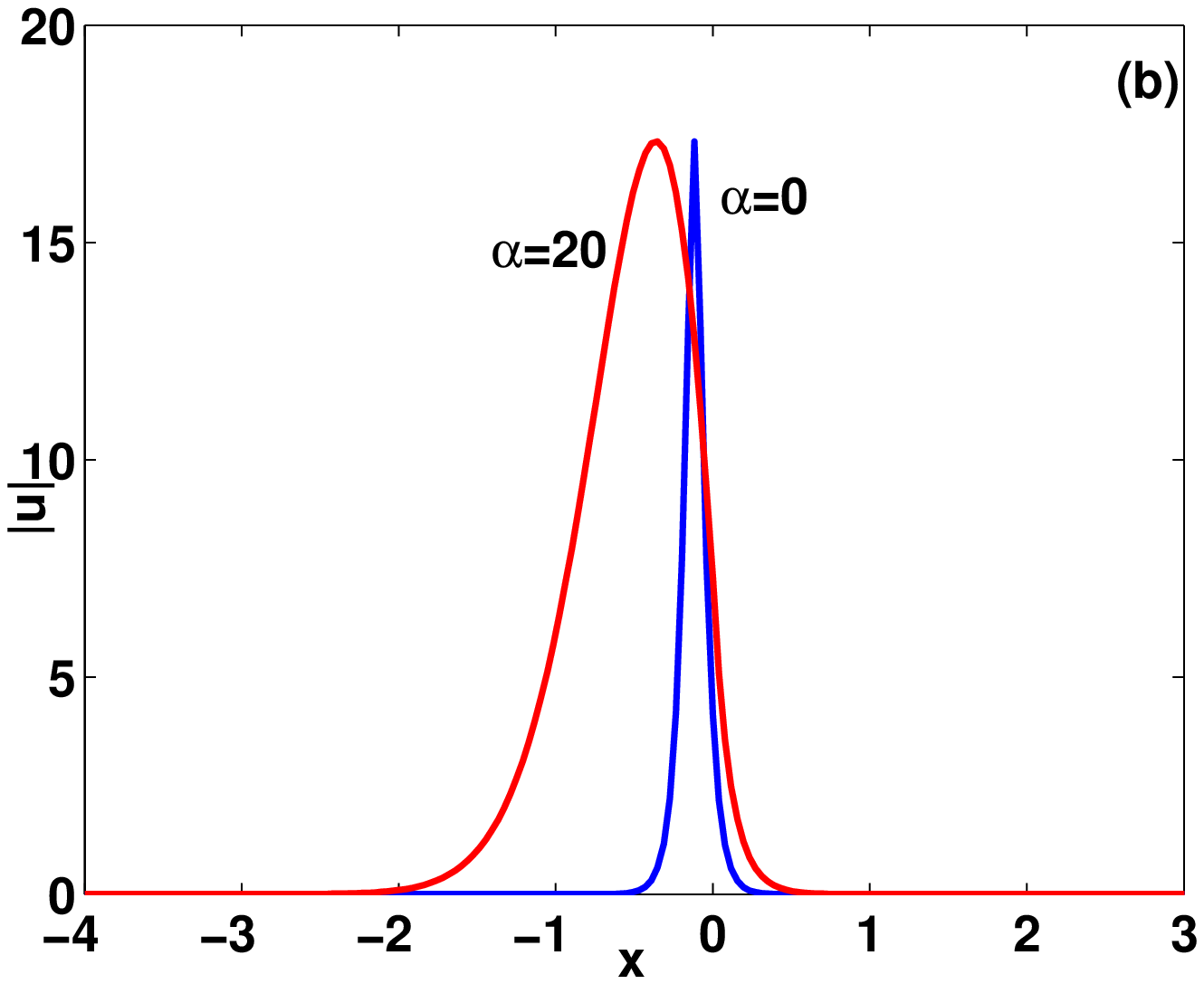}
\includegraphics[width=6.0cm]{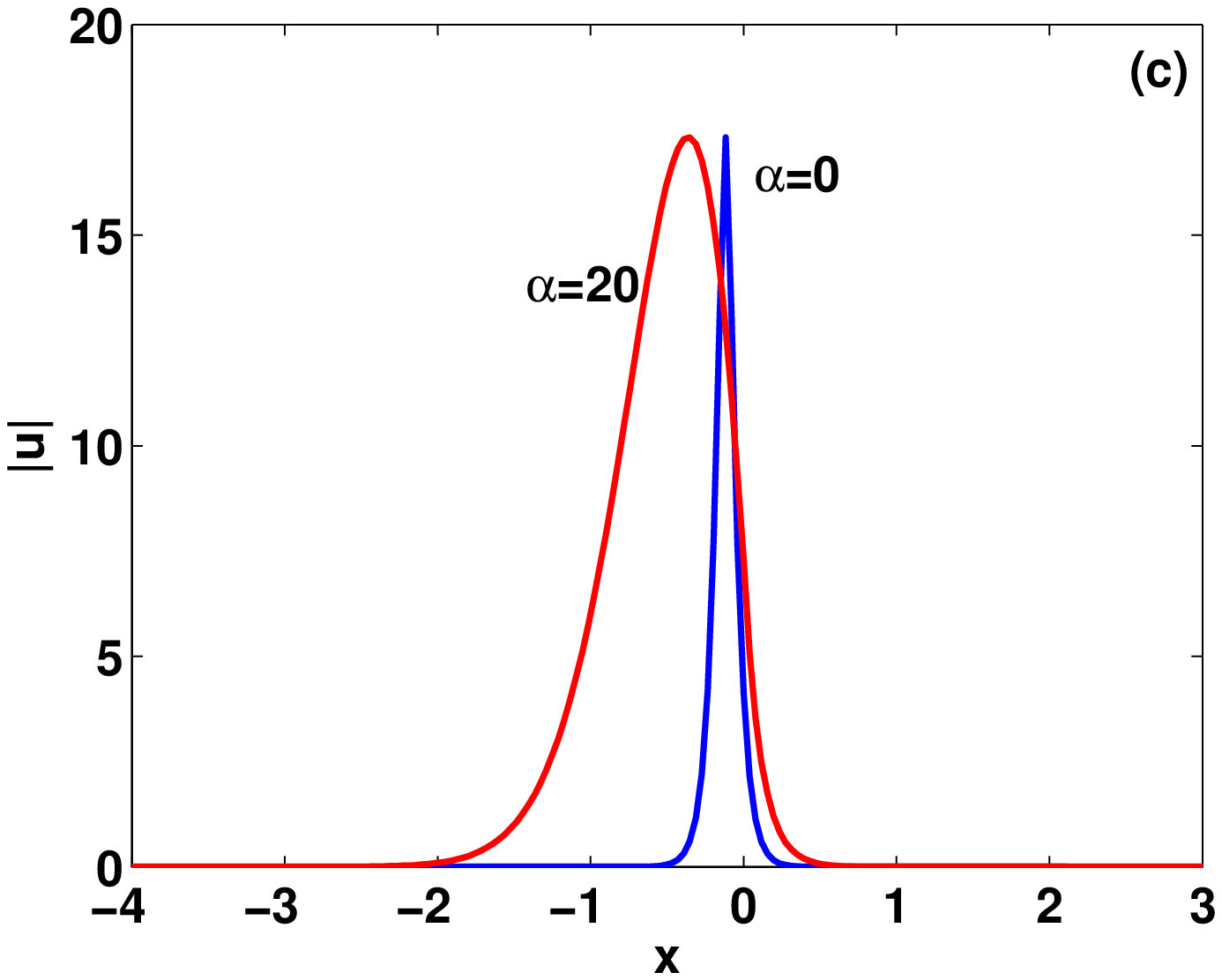}
\includegraphics[width=6.0cm]{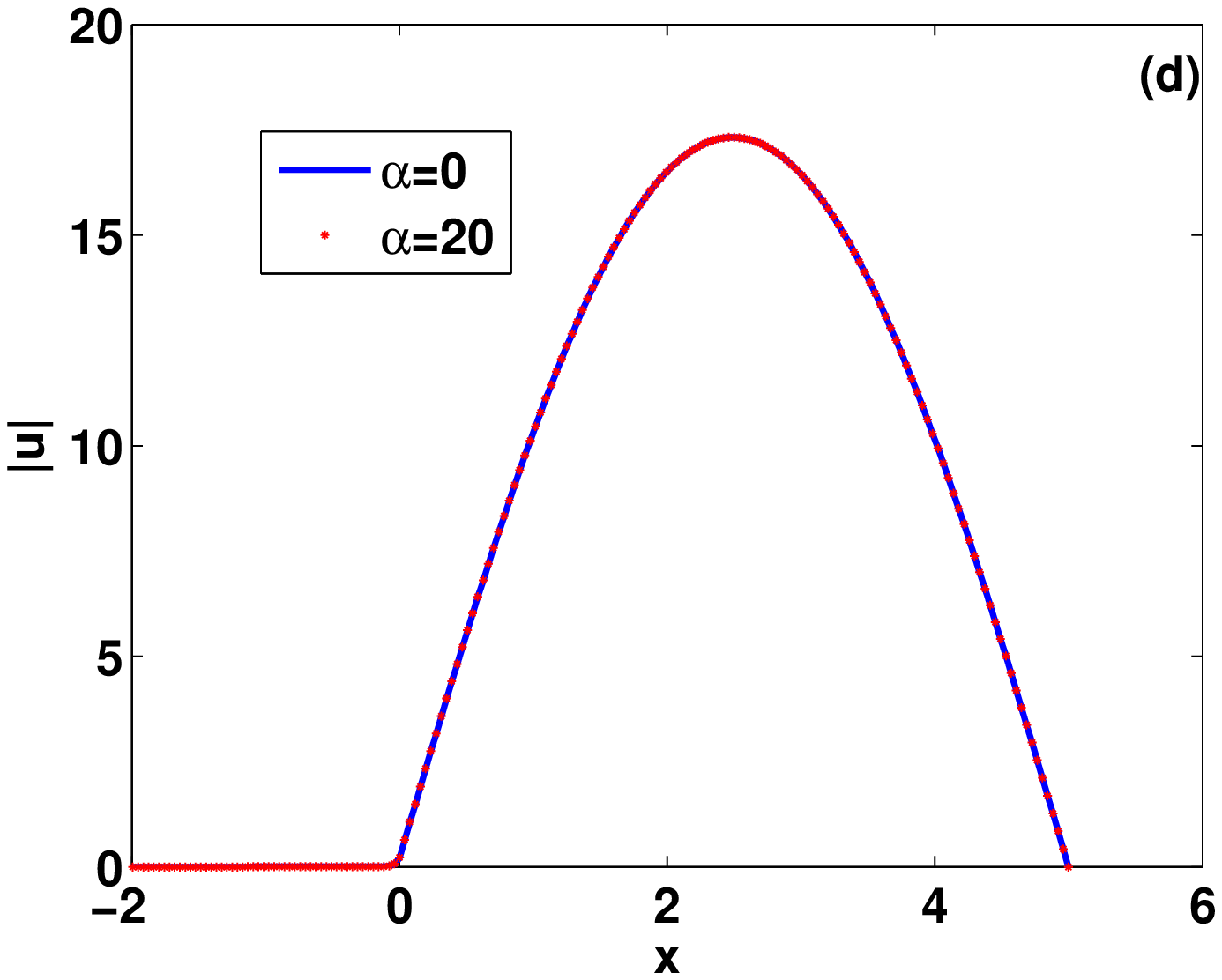}
\caption{Profiles of surface solitons at $\alpha=0$, and $\alpha=20$
for (a)$n_{L}-n_{0}=0.6$ , (b)$n_{L}-n_{0}=1\times 10^{-6}$,
(c)$n_{0}-n_{L}=1\times 10^{-6}$ and (d) $n_{0}-n_{L}=0.6$.}
\label{fig:one}
\end{figure}

\begin{figure}
\centering
\includegraphics[width=6.0cm]{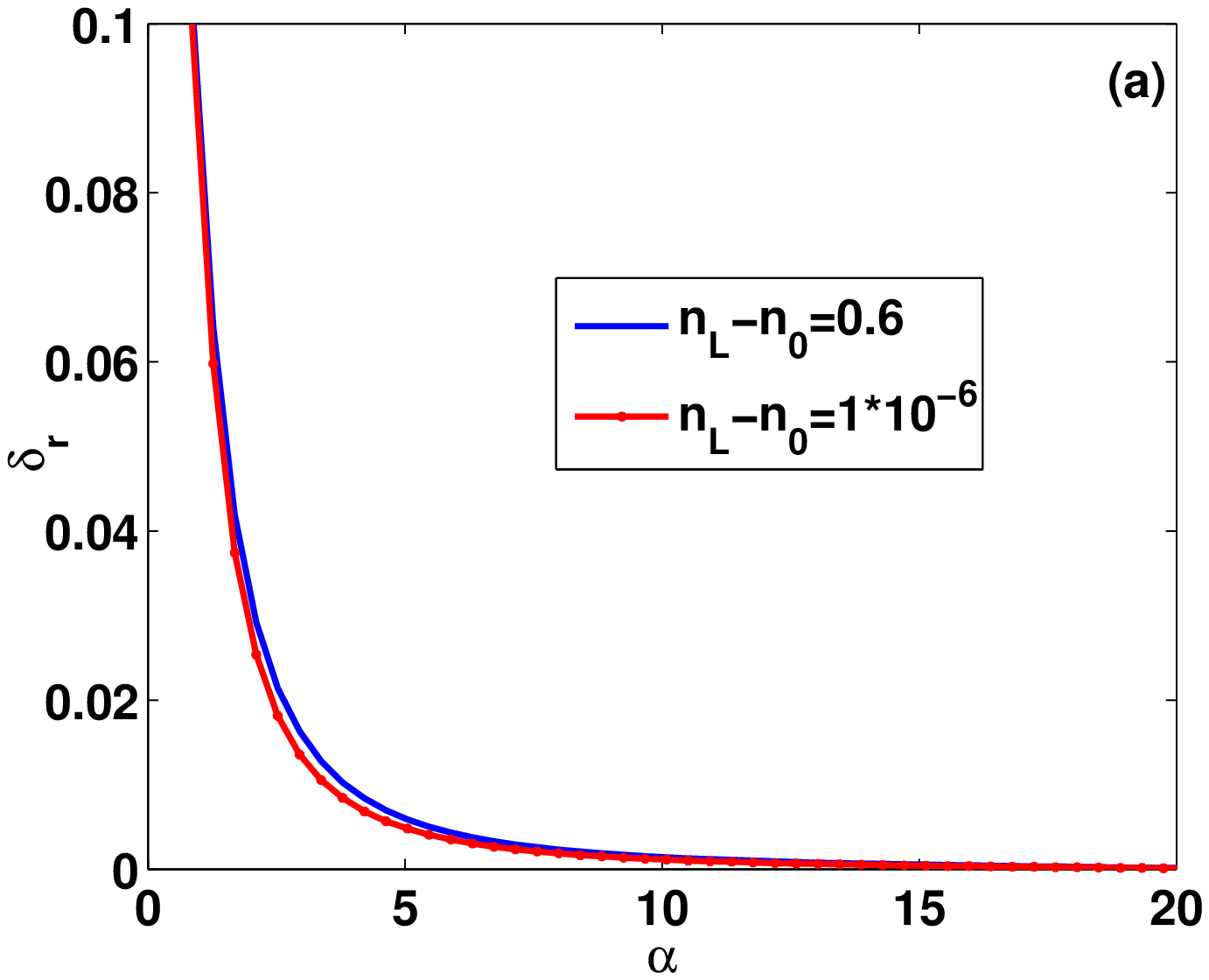}
\includegraphics[width=6.0cm]{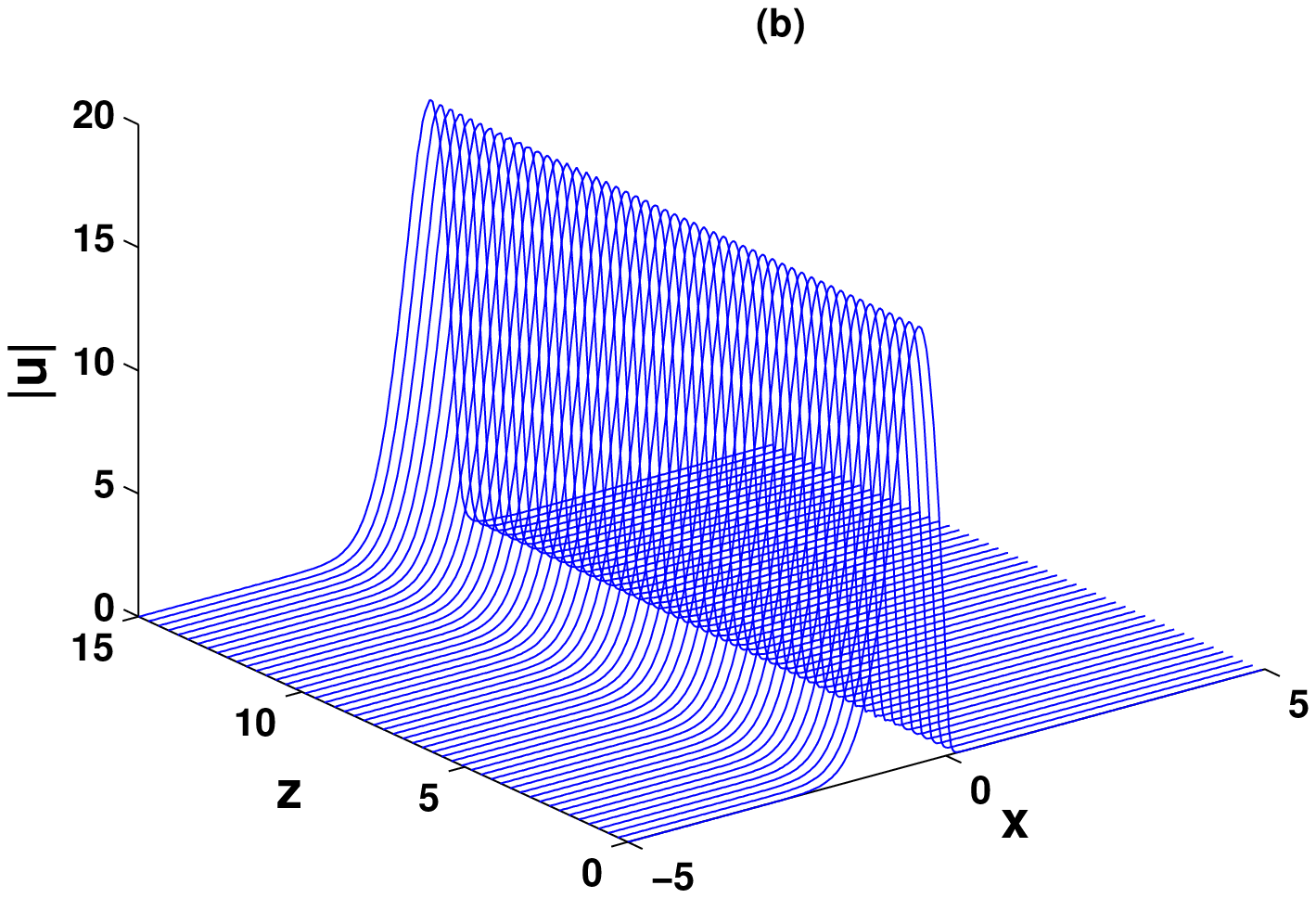}
\includegraphics[width=6.0cm]{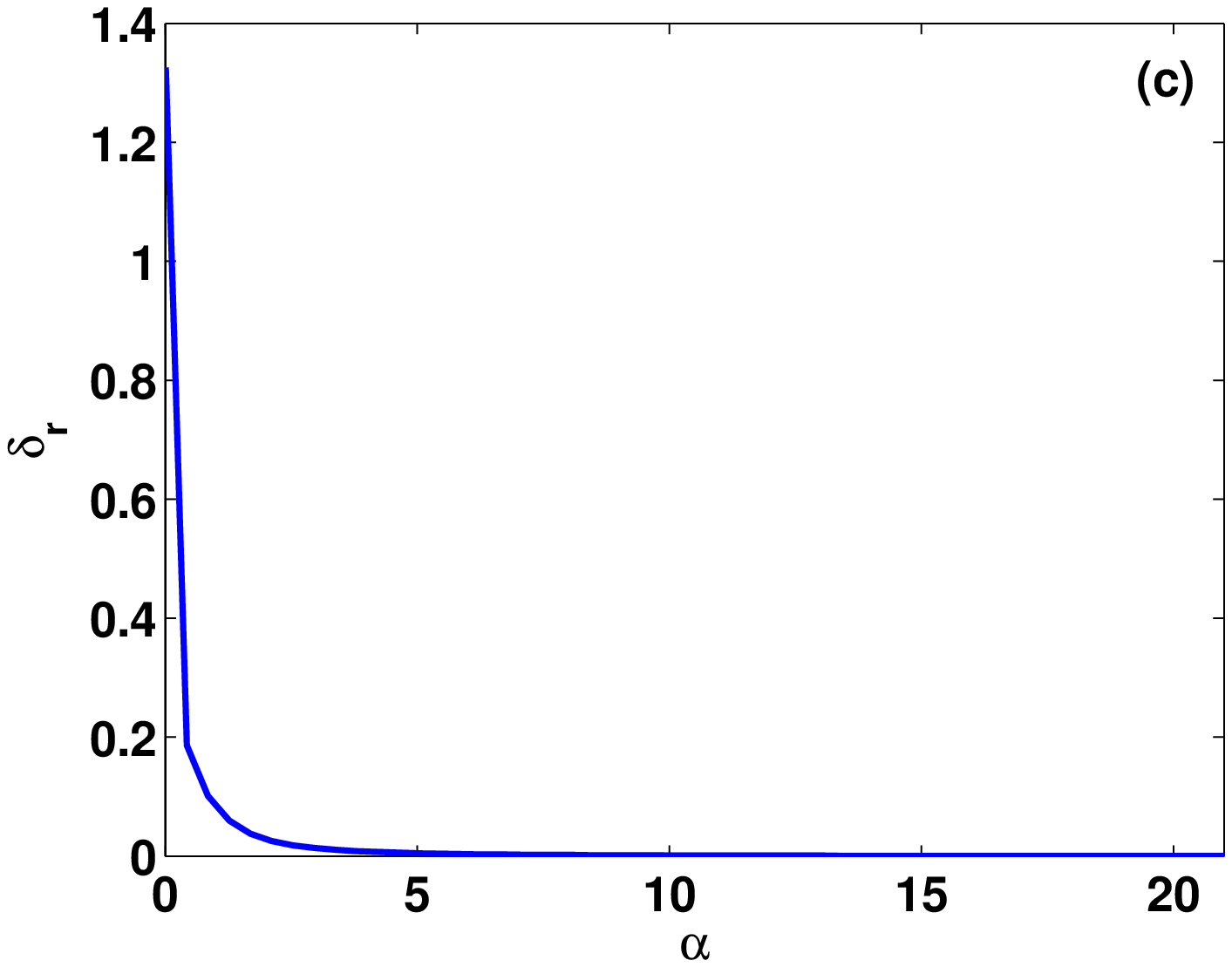}
\includegraphics[width=6.0cm]{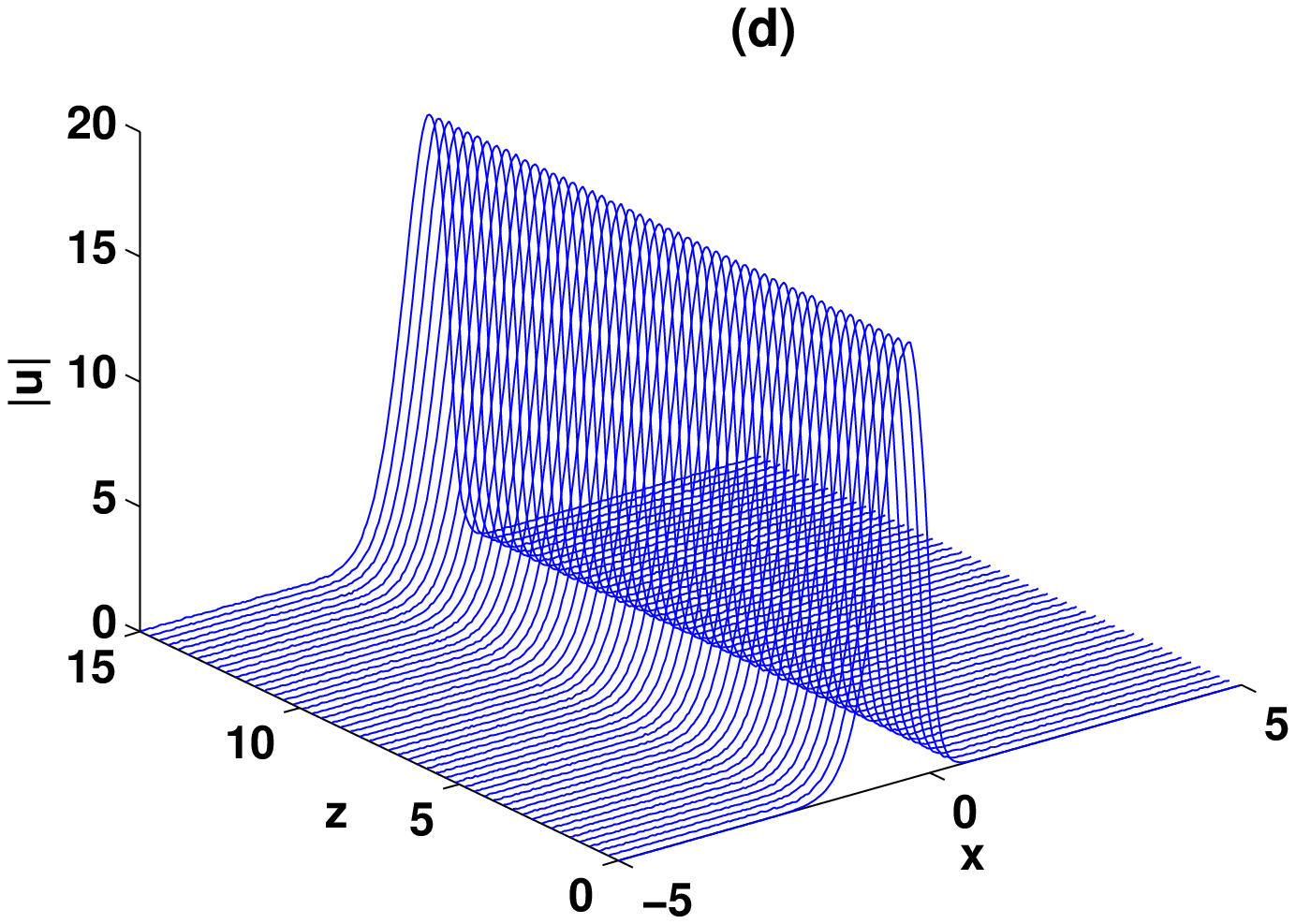}
\caption{The influence of the degree of nonlocality $\alpha$ on the
perturbation growth rate $\delta_{r}$ for $n_{L}-n_{0}=0.6$ or
$n_{L}-n_{0}=1\times 10^{-6}$(a) and $n_{0}-n_{L}=1\times
10^{-6}$(c). Propagation of the surface solitons launched at $x=0$
with 5\% noise at $\alpha=20$ for $n_{L}-n_{0}=0.6$(b) and
$n_{0}-n_{L}=1\times 10^{-6}$(d).} \label{fig:two}
\end{figure}

We first consider that the influence of the difference of $n_{L}$
and $n_{0}$ on the surface solitons. For the zero field refractive
index of the nonlinear medium($n_{L}$) is larger than the refractive
index of the linear medium($n_{0}$), at the same degree of
nonlocality of the nonlinear response, from Fig.~\ref{fig:one}, we
can find that the solitons reside almost fully inside the nonlocal
nolinear region and only weakly penetrate into the linear region
when two media have a large refractive index difference, but the
surface solitons have a significant part of their optical power
residing in the linear medium when the boundary is between two media
with a small refractive index difference which is comparable to the
nonlinear index change. The results show that the refractive index
difference affects the power distribution of the surface solitons in
two media. Comparing Fig.~\ref{fig:one}(b) with (c), we can see that
the profiles of surface solitons are alike when the refractive index
difference between two media is small. However, when the refractive
index difference between two media is big(Fig.~\ref{fig:one}(a) and
(d)), the profiles of solitons are very different and solitons are
no longer affected by nonlocality shown in
Fig.~\ref{fig:two}(d)($n_{0}-n_{L}=0.6$). Of course, when the degree
of nonlocality $\alpha$ is equal to zero, that is to say, the
nonlinear media is local, the solitons are stable in the case of
Fig.~\ref{fig:one}(c), whereas the solitons are unstable in the case
of Fig.~\ref{fig:one}(b)~\cite{ref1,ref6}.

\begin{figure}
\centering
\includegraphics[width=6.0cm]{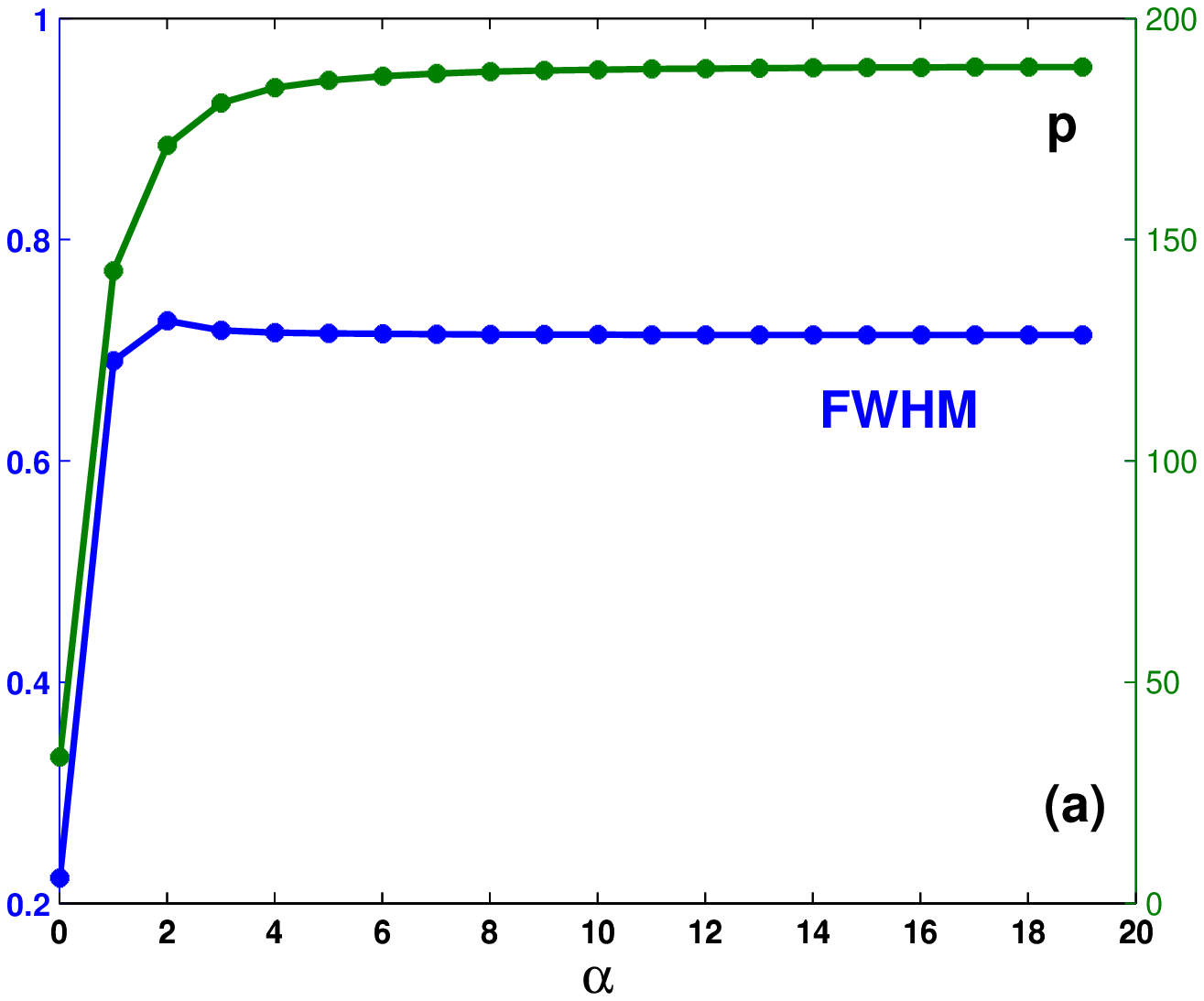}
\includegraphics[width=6.0cm]{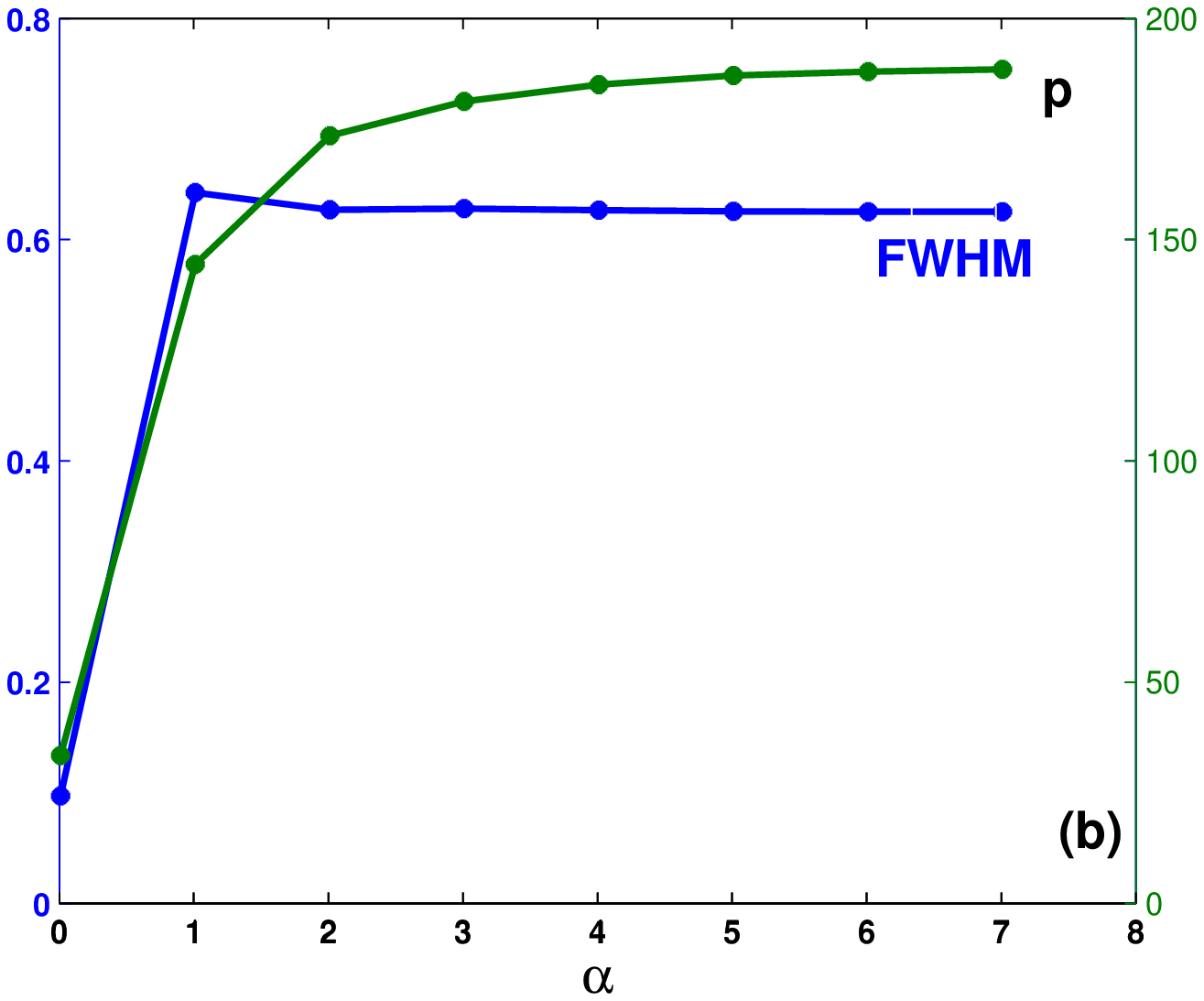}
\caption{Energy flow $P$ and FWHM versus $\alpha$ for surface
solitons for $n_{L}-n_{0}=0.6$(a) and $n_{0}-n_{L}=1\times
10^{-6}$(b).} \label{fig:three}
\end{figure}

\begin{figure}
\centering
\includegraphics[width=6.0cm]{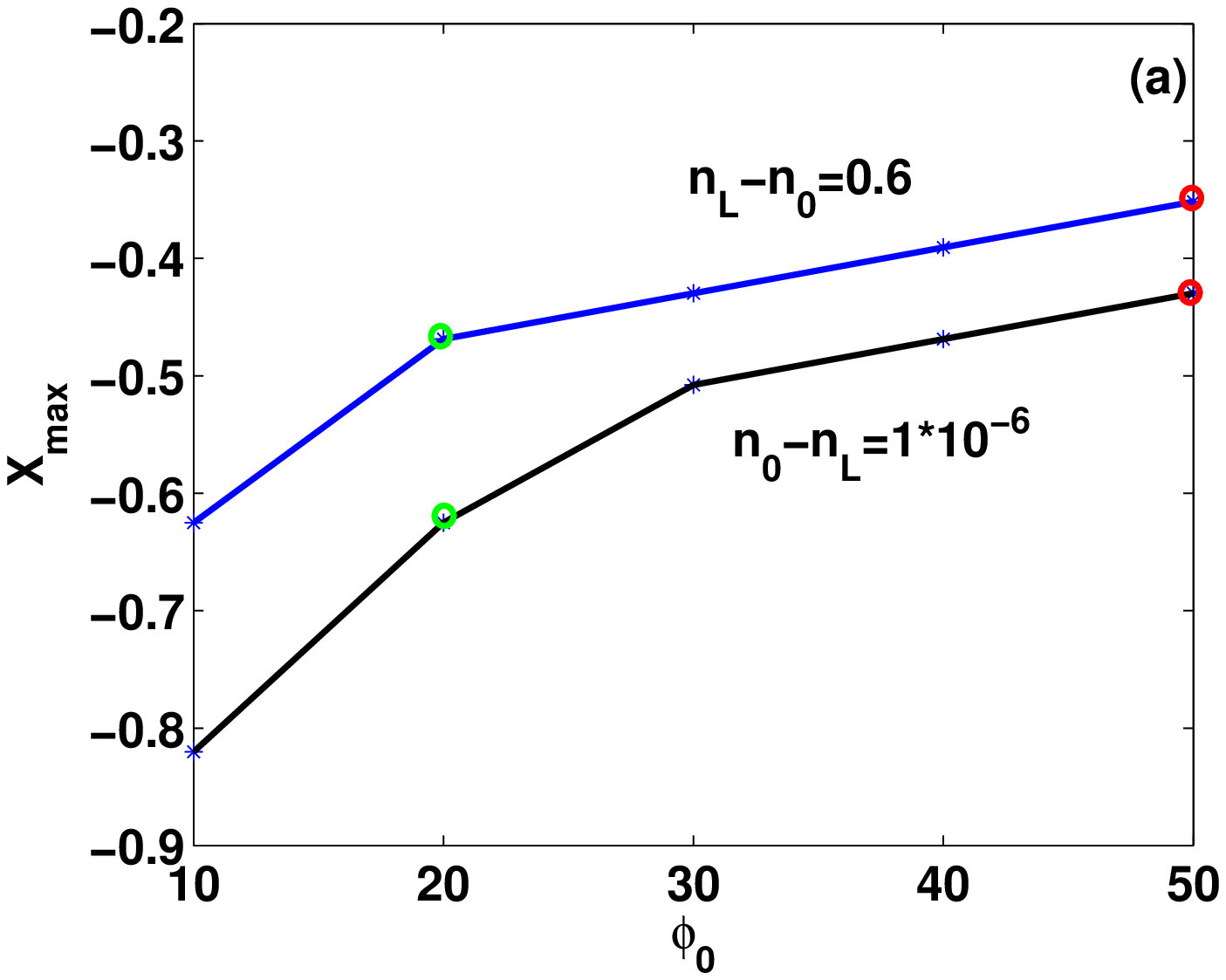}
\includegraphics[width=6.0cm]{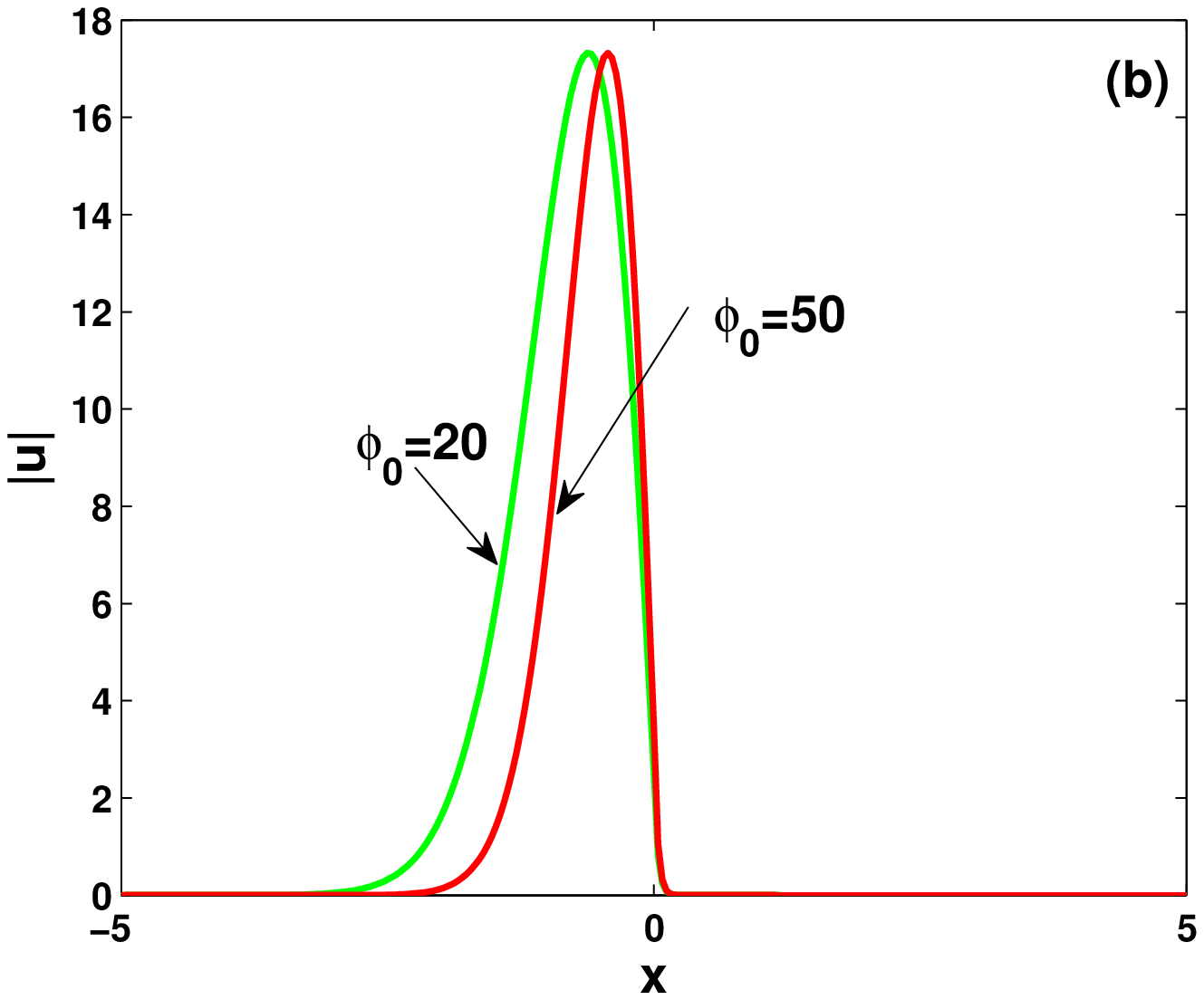}
\includegraphics[width=6.0cm]{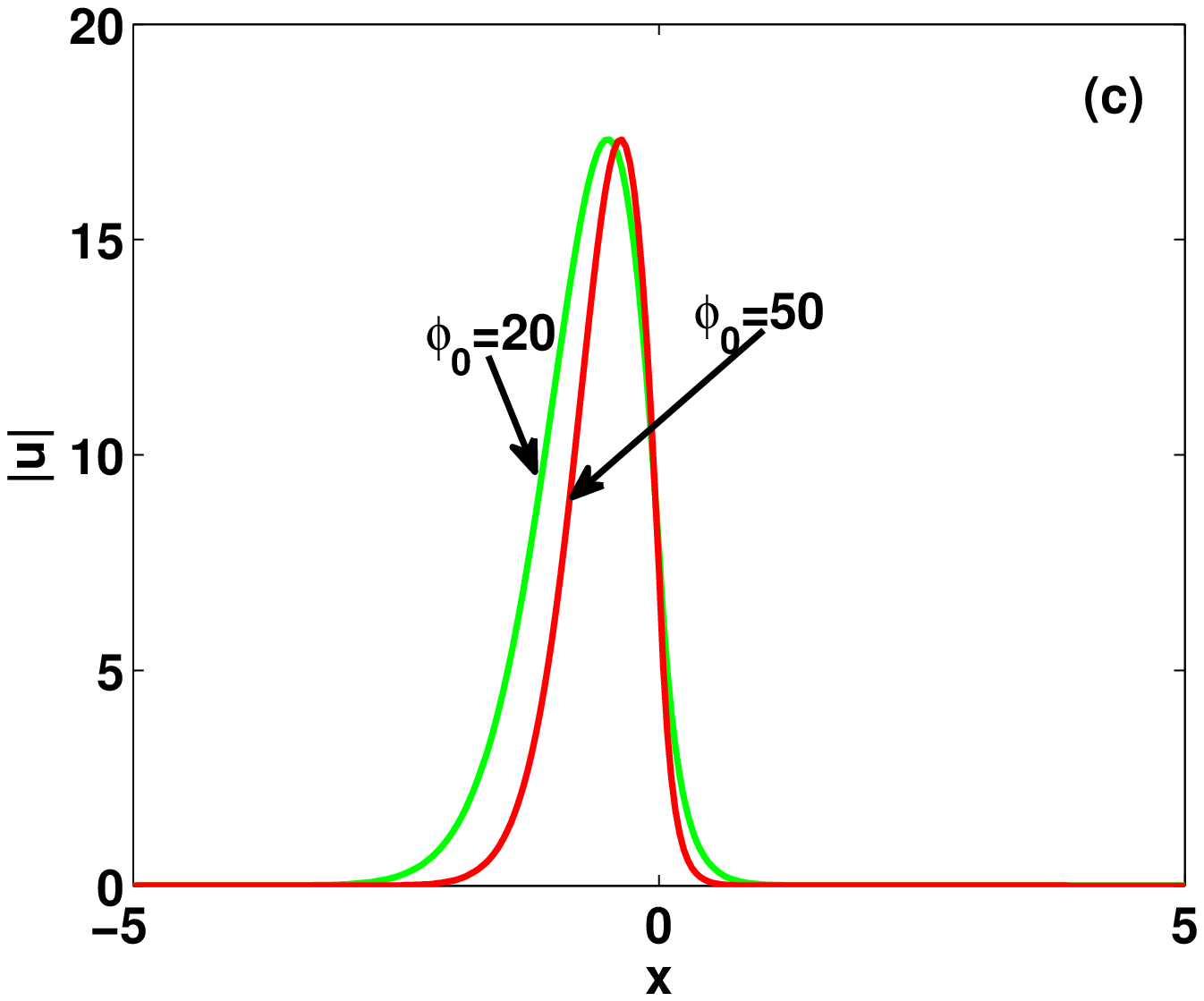}
\includegraphics[width=6.0cm]{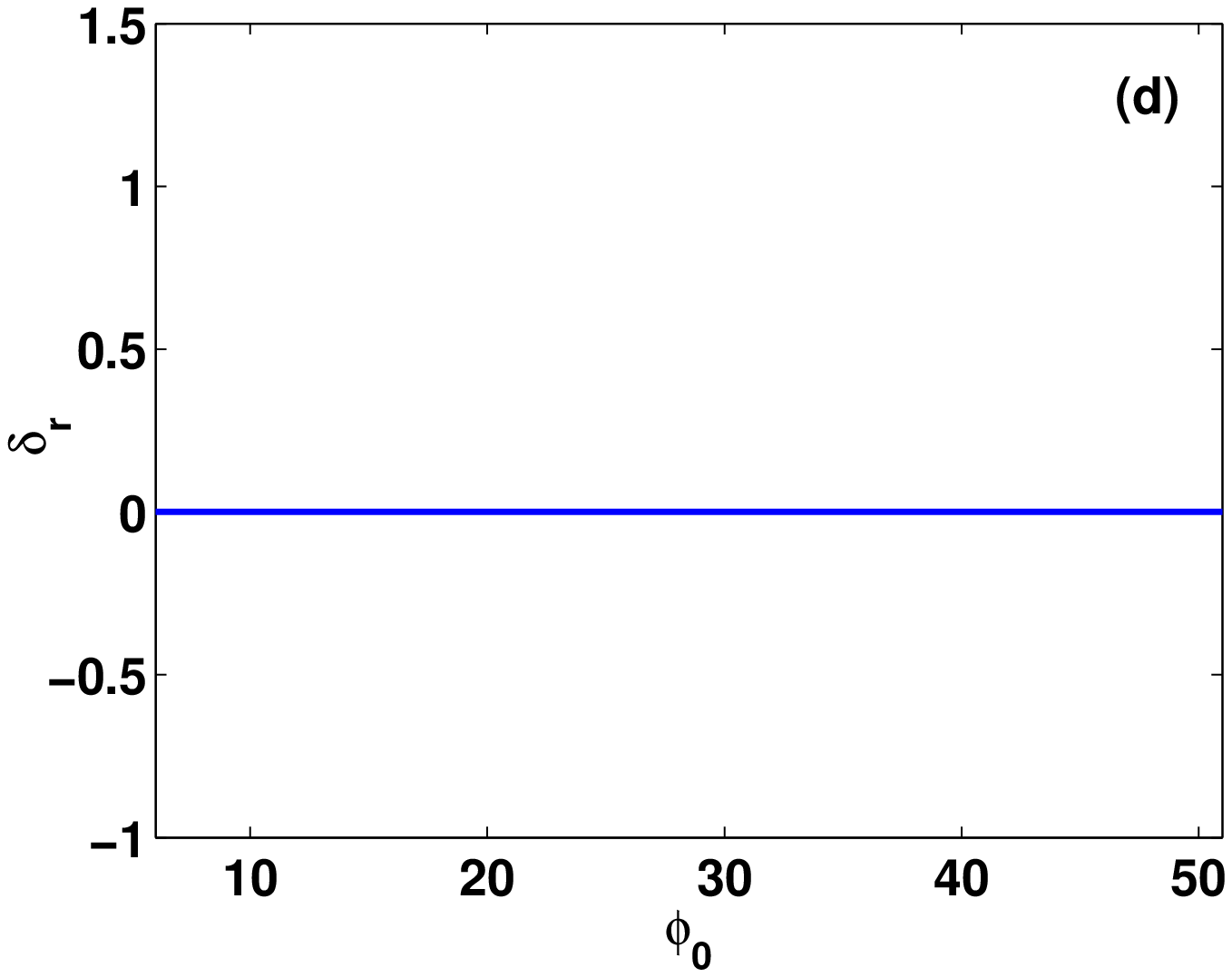}

\caption{(a)The position $X_{max}$ of the maximum value of $|u|$ as
a function of the boundary value $\phi_{0}$ at the surface. Circles
correspond to surface soltion at $\phi_{0}=20$ or $\phi_{0}=50$
shown in (b) for $n_{L}-n_{0}=0.6$ and (c) for $n_{0}-n_{L}=1\times
10^{-6}$. (d) The influence of $\phi_{0}$ on $\delta_{r}$.
$\alpha=20$ for all figures.} \label{fig:four}
\end{figure}

The central finding in this Letter is the influence of the change of
nonlocal degree on the solitons stability. In Fig.~\ref{fig:two}(a),
for $n_{L}>n_{0}$, with the degree of nonlocality becomes stronger,
the solitons are more stable. When the degree of nonlocality exceeds
a certain value, the solitons will be stable. The index difference
influences the value. For $n_{L}<n_{0}$, only when the index
difference is small, the solitons stability will be affected by the
degree of nonlocality. This is shown in Fig.~\ref{fig:two}(c).
Fig.~\ref{fig:two}(b) and (d) depict the solitons are very stable,
propagating without distortion or deviations in their trajectories
for a propagation distance of $15$ diffraction lengths with 5\%
white noise. These results illustrate the fact that the nonlocal
nonlinearity does action on the surface soltions. when the force
exerted on the beam by the nonlocal nonlinearity is equal to the
force exerted by the boundary at the interface, the solitons keep
their straight line trajectories. Here, we only show that the cases
$n_{L}-n_{0}=0.6$[Fig.~\ref{fig:two}(b)] and $n_{0}-n_{L}=1\times
10^{-6}$[Fig.~\ref{fig:two}(d)]. Having demonstrated the influence
of the degree of nonlocality on stability of the surface solitons,
we proceed to study the energy flow
$P=\int^{\infty}_{-\infty}|a|^2dX$ or FWHM of the surface solitons
as a function of the degree of nonlocality
$\alpha$[Fig.~\ref{fig:three}]. As the degree of nonlocality
increase, the energy flow monotonously increases. FWHM firstly
increases with the increase of $\alpha$, but it will decrease when
the degree of nonlocality is strongly nonlocal. Importantly, the
boundary value at the interface can also dramatically modify the
properties of surface soltions. For example, it can affect the
position $X_{max}$ of the maximum value of $|u|$
[Fig.~\ref{fig:four}(a)]. $X_{max}$ is located farther away from the
interface when the boundary value $\phi_{0}$ is smaller. In
Fig.~\ref{fig:four}(b) and (c), one can easily find this point by
comparing the surface soltion at $\phi_{0}=20$ with the surface
soltion at $\phi_{0}=50$. So, we can say that the force exerted on
the surface solitons by the interface will increase when the
boundary value increases. The force attracts the surface solitons to
the interface. However, the boundary value at the interface can not
influence the stability of solitons when $\alpha$ is a certain
value. This can be explained by Fig.~\ref{fig:four}(d) in which the
change of the perturbation growth rate $\delta_{r}$ followed by
$\phi_{0}$ is a straight line.

To summary, the stability, energy flow and FWHM of the surface wave
solitons can be affected by the degree of nonlocality of the
nonlinear medium. We find that the refractive index difference
affects the power distribution of the surface solitons in two media.
We state that the different boundary values at the interface can
lead to the different peak position of the surface solitons, but it
can not influence the solitons stability with a certain degree of
nonlocality.

This research was supported by the Specialized Research Fund for the
Doctoral Program of Higher Education (Grant No. 20060574006), and
Program for Innovative Research Team of the Higher Education in
Guangdong (Grant No. 06CXTD005).

\newpage 

\clearpage

\clearpage

\clearpage

\clearpage

\clearpage

\end{document}